\documentclass[fleqn,usenatbib]{mnras}
\usepackage{mathptmx}
\usepackage{txfonts}
\usepackage[T1]{fontenc}
\usepackage{bm}

\usepackage{savesym}
\savesymbol{iint}
\savesymbol{iiint}
\savesymbol{iiiint}
\savesymbol{idotsint}
\usepackage{amsmath}
\usepackage{graphicx}

\newcommand{\omegaupdown}{\omega_{\pm}}
\renewcommand{\omegaup}{\omega_{+}}
\newcommand{\omegadown}{\omega_{-}}

\newcommand{\nuupdown}{\nu_{\pm}}
\renewcommand{\nuup}{\nu_{+}}
\newcommand{\nudown}{\nu_{-}}

\newcommand{\nuzeroupdown}{\nu_{0\pm}}
\newcommand{\nuzeroup}{\nu_{0+}}
\newcommand{\nuzerodown}{\nu_{0-}}

\newcommand{\Bzeroupdown}{B_{0\pm}}

\newcommand{\Bzerodown}{B_{0-}}
\newcommand{\Bupdown}{B_{\pm}}
\newcommand{\Bup}{B_{+}}
\newcommand{\Bdown}{B_{-}}

\newcommand{\gammamaxup}{\gamma_{{\rm max}+}}
\newcommand{\gammamaxupdown}{\gamma_{{\rm max}\pm}}
\newcommand{\gammamaxdown}{\gamma_{{\rm max}-}}
\newcommand{\gammacut}{\gamma_{\rm c}(t)}

\newcommand{\deltatup}{\Delta t_{+}}
\newcommand{\deltatdown}{\Delta t_{-}}

\begin{document}

\title{Prospects for ultra-high-energy particle acceleration at relativistic shocks}

\author[Zhi-Qiu Huang, Brian Reville, John G. Kirk, Gwenael Giacinti]{
  Zhi-Qiu Huang,\(^{1}\) Brian Reville,\(^{1}\) John G. Kirk,\(^{1}\)
  Gwenael Giacinti,\(^{2,3,1}\)
  \\
  \(^{1}\)Max-Planck-Institut f\"ur Kernphysik, Postfach 10~39~80, 69029 Heidelberg, Germany\\
  \(^{2}\)Tsung-Dao Lee Institute, Shanghai Jiao Tong University, Shanghai 201210, P. R. China\\
  \(^{3}\)School of Physics and Astronomy, Shanghai Jiao Tong
  University, Shanghai 200240, P. R. China}

\maketitle
\begin{abstract}
We study the acceleration of charged particles by ultra-relativistic shocks 
using test-particle Monte-Carlo simulations. 
Two field configurations are considered: (i) shocks with uniform upstream magnetic field in the plane of the shock, and (ii) shocks in which the upstream magnetic field has a cylindrical geometry. Particles are assumed to diffuse in angle due to frequent non-resonant scattering on small-scale fields. 
The steady-state distribution of particles' Lorentz factors is shown to approximately satisfy \(dN/d\gamma \propto \gamma^{-2.2}\) provided the particle motion is scattering dominated on at least one side of the shock. For scattering dominated transport, the acceleration rate scales as \(t_{\rm acc}\propto t^{1/2}\), though recovers Bohm scaling \(t_{\rm acc}\propto t\) if particles become magnetised on one side of the shock. For uniform field configurations, a limiting energy is reached when particles are magnetised on both sides of the shock. For the cylindrical field configuration, this limit does not apply, and particles of one sign of charge will experience a curvature drift that redirects particles upstream. For the non-resonant scattering model considered, these particles preferentially escape only when they reach the confinement limit determined by the finite system size, and the distribution approaches the escapeless limit \(dN/d\gamma \propto \gamma^{-1}\). The cylindrical field configuration resembles that expected for jets launched by the Blandford \& Znajek mechanism, the luminous jets of AGN and GRBs thus provide favourable sites for the production of ultra-high energy cosmic rays. 
\end{abstract}


\section{Introduction}
Relativistic shocks occur in many astrophysical sources of non-thermal
emission such as pulsars, gamma-ray bursts (GRBs), micro-quasars and
active galactic nuclei (AGN) and there is a growing wealth of
observational evidence indicating that these shocks efficiently
convert a large fraction of the energy they process into extremely
energetic particles. Acceleration by the first-order Fermi mechanism
operating at shocks has been proposed and thoroughly investigated
by numerical and analytical approaches for the case of parallel shocks.
However, 
interpreting recent observations, in particular those at very high
energy
\citep{2019Natur.575..459M,2019Natur.575..464A,2021Sci...372.1081H},
requires a clear understanding of, at the very least,
the spectral index predicted and the
maximum particle energy achievable under more realistic
physical conditions. Even in the test-particle approximation, our knowledge
of these quantities remains incomplete. 

Our goal in this work is to use test-particle Monte Carlo simulations
to predict these quantities in two relevant field configurations: that of
a uniform magnetic field, and that of a cylindrically symmetric field around
a current carrying axis, such as can be expected in 
the jets that are either
directly observed or inferred in all of the classes of object listed
above. In each case we concentrate on a planar, perpendicular shock front ---
one in which
the magnetic field and the shock lie in the same plane. This
is the generic situation downstream of a highly relativistic shock front,
since the component of the field in the shock plane is compressed by
roughly the Lorentz factor of the shock, whereas the
component along the shock normal is unchanged
\citep{1990ApJ...353...66B}. 

The test-particle theory of particle acceleration at relativistic
shocks was mostly developed in the decade following the original works
outlining the non-relativistic theory \citep[see, for example,][for a
review.]{1999JPhG...25R.163K} Two complementary approaches have been
used: (i) Monte-Carlo simulation in which particles move in a simple,
prescribed magnetic field geometry whilst undergoing stochastic
transport, either represented as a sequence of small-angle scattering
events
\cite[e.~g.][]{1988A&A...201..177K,1993MNRAS.264..248O,2012ApJ...745...63S}
or, more formally, treated using stochastic differential equations
\citep{2001MNRAS.328..393A,2015ApJ...809...29T} and (ii) direct
numerical integration of particle orbits in a synthetically
constructed, turbulent magnetic field
\cite[e.~g.][]{1991MNRAS.251..438B,2006ApJ...650.1020N, 2006MNRAS.366..635L}.  Both methods have the drawback
that they rely on a poorly constrained prescription for the
turbulence. However, the former method, which we adopt here, has the
advantages that it can be benchmarked against approximate analytic
solutions, can model the energy dependent scattering expected to arise
from self-excited fluctuations, and requires relatively modest
computing resources.

Kinetic particle-in-cell (PIC) simulations do not have the fundamental
limitations inherent in the test-particle approximation and have
recently advanced towards a self-consistent study of relativistic
shocks \citep{2020Galax...8...33V,2021PhRvL.127c5101S,2023arXiv230311394B}.
In such
simulations, Fermi acceleration is routinely observed as a consequence
of shock formation, at least for those shocks in which the upstream
medium is sufficiently weakly magnetized. These simulations provide
valuable insights into the physical processes that initiate
acceleration, and, although they are currently limited to a narrow
range of energy and length scales, nevertheless motivate the form of
the scattering operator used in our analytical and Monte-Carlo work.

In the light of GRB afterglow detections at TeV gamma-ray energies, \citet{2022ApJ...925..182H}
used these insights to place constraints on the maximum
electron energy expected at an ultra-relativistic shock.  This revealed tension
between observations and the simple one-zone synchrotron self-Compton
emission model.  A partial resolution was proposed in a
companion paper \citep{Kirketal23}, which used both
analytical and Monte-Carlo methods to show that a strong, uniform
downstream magnetic field does not necessarily inhibit acceleration, as
had previously been supposed \citep{2001MNRAS.328..393A}.
Here, we extend these studies in section \ref{uniformmagneticfield},
using the Monte-Carlo method to
quantify the maximum energy reached at a shock front
in a uniform magnetic field scenario.

While the uniform field configuration is instructive and can be applied
in many cases, it is inadequate when
the length scales associated with the highest energy particles become
comparable to those of preexisting structure in the magnetic field.
One interesting example is that of the termination shock of a pulsar
wind, since particles accelerated at latitudes within one gyroradius of the
equatorial current sheet can experience a reversal of the magnetic
field during their excursions both upstream and downstream. This
situation was treated using the synthetic field method by \citet{GK18}
and global PIC simulations were performed by \citet{CeruttiGiacinti}.
In each case, spectra harder than those predicted for relativistic shocks in a homogeneous magnetic field were observed \citep[see
also][]{2013ApJ...770..147C, 2019MNRAS.487..952C}.

Motivated by this finding, and by the fact that jets are a common feature in
the sources of interest, we consider, in section \ref{cylinder},
the related configuration
of a 
cylindrically symmetric magnetic field. This is a reasonable model
of the field upstream of the termination shock of a current-carrying jet, or
upstream of the 
forward shock of an explosion that propagates along the rotation axis
of a progenitor star into its magnetized wind.
We identify a spectral
break from the standard result at lower energy to one with
a harder index
at higher energy. As in the case addressed by \citet{GK18}, the new component
consists of particles of only one sign of charge. We find that the
harder spectral index is almost
independent of the level of scattering and that this component can
extend to extremely high energy, limited only by the transverse size
of the jet.

The paper is organized as follows. In Section \ref{code}, we motivate
the transport model that we implement in our Monte-Carlo
method. Numerical results for the uniform and cylindrical field
cases are presented in Sections \ref{uniformmagneticfield} and
\ref{cylinder}, respectively, and the implications and limitations of our
results are discussed in Section \ref{discussion}.

\begin{figure*}
\begin{center}
\includegraphics[width=0.45\textwidth]{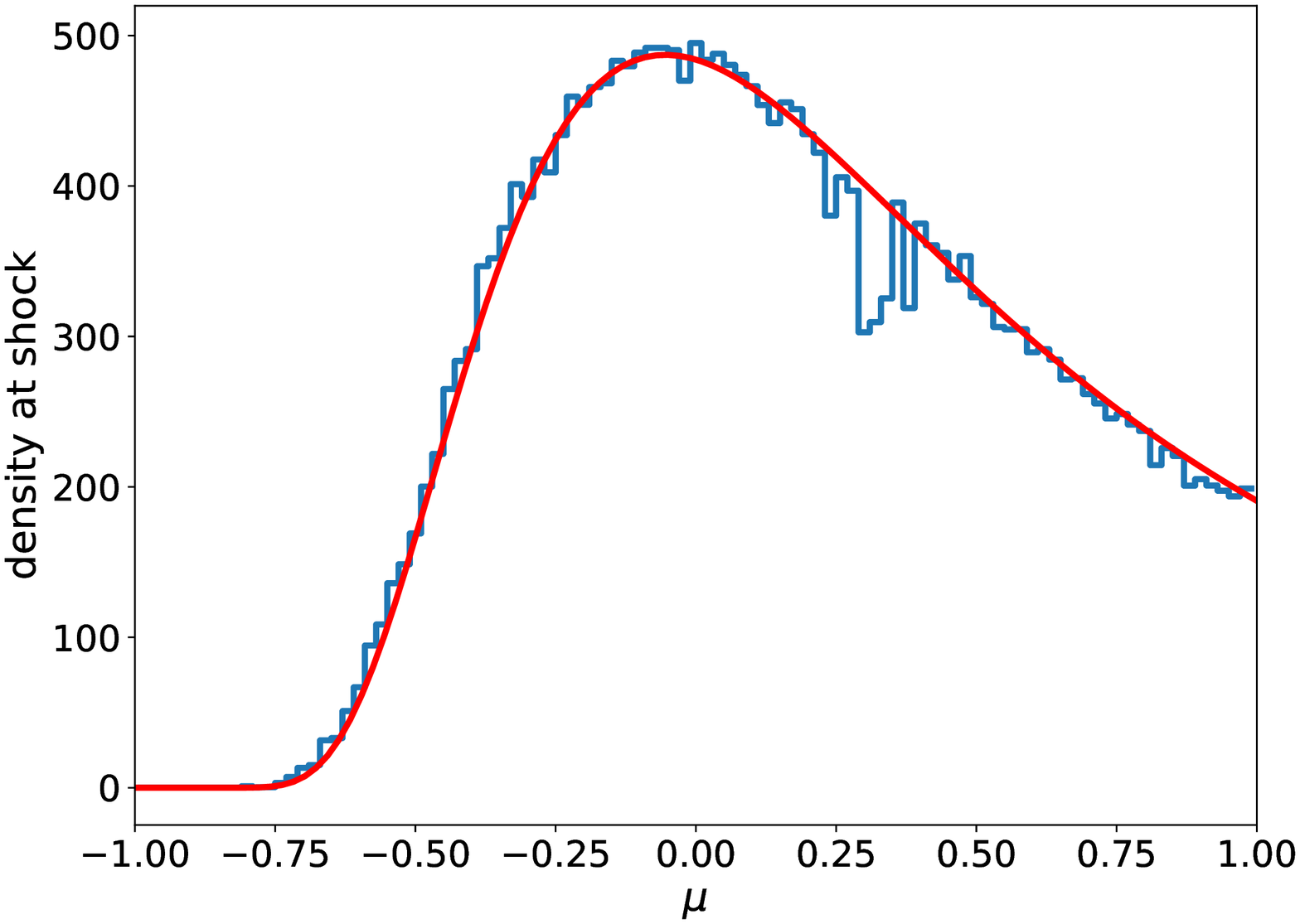}
\includegraphics[width=0.45\textwidth]{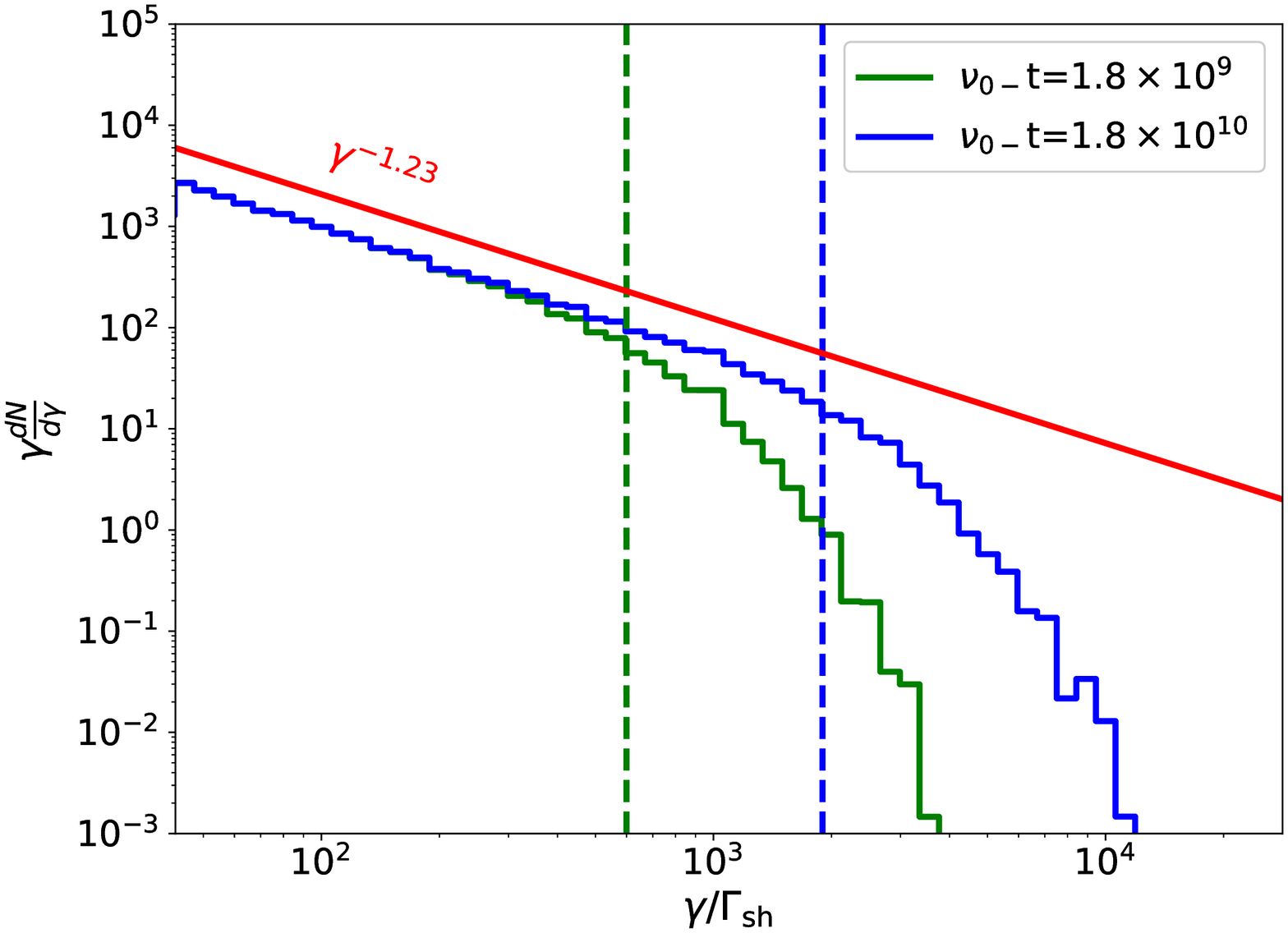}
\caption{Results for the case of negligible large-scale magnetic
  field (equivalent to the parallel shock case),
  shown in the downstream frame. Left panel: The dependence
  of the phase space density \(f\) at the shock front on the cosine
  \(\mu\) of the angle between
  \(\bm{p}\) and the shock normal
  (blue) compared to the analytic approximation. Right panel: the
  dependence of \(\gamma\textrm{d}N/\textrm{d}\gamma = \gamma^3 f\) at
  the shock front on the normalized particle Lorentz factor
  \(\gamma/\Gamma_{\rm sh}\) at two different times, compared to the
  predicted power-law (red line) and the estimated maximum energy at
  the chosen times (green and blue vertical
  lines, see eq~(\ref{eq:maxgammaestimate}))}.
    \label{fig:pure_scattering}
\end{center}
\end{figure*}

\section{Test-particle transport}
\label{code}

The distribution function  \(f(t, \bm{x}, \bm{p})\) in  phase space
\((\bm{x},\bm{p})\) of test particles
that undergo small-angle, elastic scatterings satisfies the Fokker-Planck equation
\begin{equation}
\frac{\partial f}{\partial t} + \dot{\bm{x}}\cdot\frac{\partial f}{\partial \bm{x}} + \dot{\bm{p}} \cdot \frac{\partial f}{\partial \bm{p}} =
\frac{\nuupdown}{2}\Delta_{\bm{p}} f\,,
\label{fpeq}
\end{equation}
where \(\Delta_{\bm{p}}\) is the angular part of the Laplacian in
momentum space, \(\nuupdown\) the isotropic scattering rate (that
can depend on \(\bm{x}\) and \(p\))  and
\((\dot{\bm{x}},\dot{\bm{p}})\) are the time derivatives of an
unscattered trajectory in phase space. Equation (\ref{fpeq}) applies
in a reference frame, called the local fluid frame,
in which the electric field vanishes. The unscattered trajectories of a
particle of charge \(q\) and mass \(m\) are then determined 
by the large-scale, static magnetic field \(\bm{B}\), i.~e.,
\(\dot{\bm{p}} = q(\dot{\bm{x}}\times\bm{B})/c\).
In the presence of a shock front that lies in the plane \(x=0\) in cartesian
coordinates, the two fluid frames of interest are the 
upstream \(x>0\), suffix
\lq\lq\(+\)\rq\rq\ and downstream \(x<0\), suffix \lq\lq\(-\)\rq\rq, frames,
each with a corresponding scattering rate \(\nuupdown\).
We will assume, for simplicity, that these frames are connected by a Lorentz boost in the \(x\)-direction. 
Length and timescales in the
local frame are conveniently normalized using the non-relativistic gyrofrequency
\(\omegaupdown=\left|q\Bzeroupdown\right|/mc\)
associated with an upstream/downstream fiducial field strength \(\Bzeroupdown\).
When a trajectory reaches the shock front, it is assumed to emerge
into the region on the other side without changing its momentum. 

The nature of the particle transport and, hence, the spectrum and
angular distribution of accelerated particles, depends on the relative
importance of scattering and deflection in the large-scale field in
each half-space. For sufficiently weakly magnetized shocks, the Weibel
instability, which operates initially on the scale of the plasma
skin-depth, drives the growth of highly non-linear magnetic field
structures in the shock-transition region, extending both into the
upstream and downstream regions \cite[e.~g.][]{2013ApJ...771...54S,
  2020Galax...8...33V}.  Assuming these structures remain small
compared to the gyroradius of a test particle, they are responsible
for nonresonant scattering.  Then, following \citet{KirkReville},
for a characteristic fluctuation length scale \(\lambda\) of
volume averaged root mean square amplitude \(\delta B\),
the mean scattering
angle per fluctuation is \(\Delta \theta \approx
\left|q\right| \delta B \lambda / pc\). Taking the mean
time between scatterings to be \(\lambda /c\) one finds a
scattering rate proportional to \(p^{-2}\), characteristic of non-resonant interactions:
\begin{align}
\nuupdown &= \nuzeroupdown\left(p/mc\right)^{-2}\,,\\
             \noalign{\hbox{where}}
             \nuzeroupdown&=\left(\delta B/\Bzeroupdown\right)^2\omegaupdown^2\lambda/c
\end{align}
is independent of particle energy.

Simulations are required to determine the properties of \(\delta B\) and
\(\lambda\). Filamentary structures driven by the Weibel instability
typically develop on length-scales
\(\lambda = 10-100 c/\omega_{\rm pi}\) where
\(\omega_{\rm pi} = \sqrt{4 \pi n^2 e^2 c^2/w}\) is the relativistically
corrected plasma frequency, with \(w\) the enthalpy density. The growth
of longer wavelength fluctuations in the foreshock region was
investigated for unmagnetized \citep{2009ApJ...696.2269M} and
magnetized \citep{2014MNRAS.439.2050R} ambient plasma
conditions \cite[see also][]{2006ApJ...651..979M}. However, a complete multi-scale theory of turbulent field
generation at ultra-relativistic shocks does not exist at present, and
for simplicity, we therefore treat the scattering rate as a free
parameter that can take different values upstream or downstream, but
is otherwise homogeneous.

We employ a Monte-Carlo code --- see Appendix~\ref{appendixA} --- to
construct time-dependent solutions of Eq.~(\ref{fpeq}) upstream and
downstream of a relativistic shock front, measuring \(f\) at the shock
front and assuming an injection term that is zero for \(t<0\) and
constant for \(t>0\).

\section{Results}
\subsection{Uniform magnetic field}
\label{uniformmagneticfield}

\subsubsection{Parallel shock}
\label{sec:pure_scattering}
We first use our code to revisit the case in which the large-scale
field can be neglected. This is equivalent to the case of an exactly
parallel shock front, in which the strength of the magnetic field
plays no role.  We adopt a constant Lorentz factor
\(\Gamma_{\rm{sh}}=50\sqrt{2}\) for the shock, as seen by an upstream
observer, and a downstream shock speed of \(\beta_{\rm d}=1/3\),
as expected from the hydrodynamic
jump conditions for an ultra-relativistic shock propagating in a cold gas. This
case is not expected to differ significantly from the
ultrarelativistic limit \citep{Kirketal23}, where an analytic approximation is
known for the stationary (time-asymptotic) distribution function at
\(\gamma\gg\Gamma_{\rm sh}\) and estimates
of the acceleration rate are available, thereby enabling the code to be
benchmarked.

Fig.~\ref{fig:pure_scattering} shows the
angular distribution of particles at the shock front (left) and the
accelerated particle spectrum (right),
both measured in the downstream frame.
(In the angular distribution, we
record only those particles with energy
larger than \(30\) times that of injection, for which no memory of the
injection conditions remains. In the spectrum we plot the number of these
particles per logarithmic energy interval,
\(\gamma\textrm{d}N/\textrm{d}\gamma = \gamma^3 f\).)
For comparison, we plot, in the left panel,
the angular dependence of the
leading eigenfunction \citep{2000ApJ...542..235K}, which
shows close agreement with our simulation results. We note the 
feature associated with particles that
graze the shock, here occurring at \(\mu=1/3\).

In the right panel,
we compare the expected time-asymptotic power law spectrum with
the simulation results at two different times. 
Below a cut-off, \(\gammacut\),
that advances to higher energy with increasing time,
the distribution is found to match closely the predicted
stationary value \(\gamma\textrm{d} N/\textrm{d}\gamma \propto \gamma^{-1.23}\) 
(\(f\propto \gamma^{-4.23}\)) 
for an ultrarelativistic shock.
The time-dependent position of this cut-off,
is predicted by equating 
the energy dependent acceleration timescale with the time elapsed since
injection started \citep[see eq~(43)
in][]{2001MNRAS.328..393A}.
The acceleration timescale is simply the sum of the average time
spent upstream, \(\deltatup\), and downstream, \(\deltatdown\)
--- both measured in the downstream rest frame --- during one cycle 
(which starts at the shock, crosses it
once and then returns to it).
When scattering dominates both up and downstream,
\begin{align}
  \deltatup&=\gamma^2/\left(\Gamma_{\rm sh}\nuzeroup\right)
             \nonumber\\
  \deltatdown&=\gamma^2/\nuzerodown\,,
\end{align}
therefore,
\begin{align}
  \gammacut&=t^{1/2}
    \left(\frac{1}{{\nuzerodown}}+\frac{1}{\Gamma_{\rm sh}{\nuzeroup}}\right)^{-1/2}\,.
  \label{eq:maxgammaestimate}
\end{align} 

In Fig~\ref{fig:pure_scattering} we adopt
\(\nuzeroup = \nuzerodown\), in which case
accelerating particles spend most of their time in the downstream
region during one complete cycle. Since the shock does not decelerate and
losses and boundary effects are neglected, the cut-off increases indefinitely.
However, its rate of increase slows down
because of the quadratic dependence of the scattering time on energy
\cite[cf.][]{Stockem,2013ApJ...771...54S,Plotnikovetal18}. The time-dependent spectrum softens significantly above the Lorentz factor given by Eq.~(\ref{eq:maxgammaestimate}), shown in the figure as green and blue vertical lines.

\begin{figure}
\begin{center}
\includegraphics[scale=0.35]{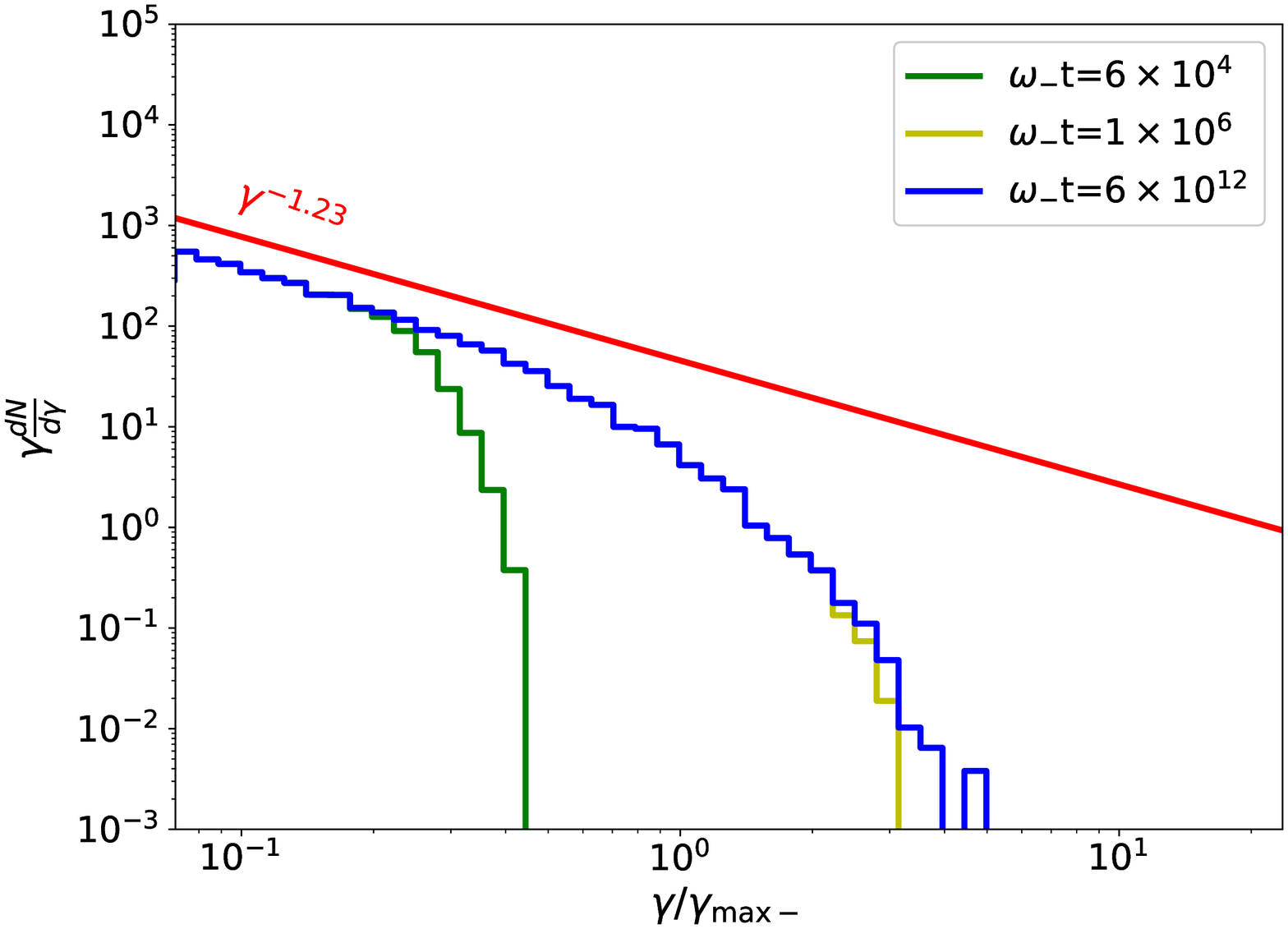}
\caption{
  The particle spectra at a perpendicular shock in a uniform magnetic field
  with relatively weak scattering upstream
\(\nuzeroup\, =\,10^{-4}\nuzerodown\), as a function of the ratio of the Lorentz factor to its predicted maximum, \(\gammamaxdown
={\nuzerodown}/\omegadown=3\times10^4\), see eq~(\ref{eq:dmldefinition}).
The upstream magnetized limit, \(\gammamaxup\approx 600\),
lies close to the injection energy and is not shown. 
  \label{fig:spectrum_weakest}}
\end{center}
\end{figure}

\subsubsection{Perpendicular shock}
\label{perpendicular}

We next consider the case where the regular magnetic field is
perpendicular to the shock normal. This {\em perpendicular} shock is
the generic configuration for relativistic shocks, because the
velocity of the point of intersection of a magnetic field line and the
shock surface exceeds \(c\) unless the shock normal happens to be
aligned with the upstream magnetic field to within an angle of
\(1/\Gamma_{\rm sh}\).  (Note that all such {\em superluminal shocks}
can be transformed into perpendicular shocks by a Lorentz boost along
the shock surface.)  The downstream magnetic field is determined by
the ideal magneto-hydrodynamic shock jump conditions for a weakly
magnetized ultra-relativistic shock i.~e.,
\(\Bdown\, =\, 2\sqrt{2}\Gamma_{\rm{sh}}\Bup\)
\cite[e.~g.][]{1999JPhG...25R.163K}, where \(\Bupdown\) are measured
in the corresponding rest frame of the plasma. Since the fields up and downstream are uniform, they can be taken to equal their fiducial values
\(\Bupdown\equiv\Bzeroupdown\). We again adopt \(\Gamma_{\rm sh}=50\sqrt{2}\)
and set \(\nuzerodown=3\times10^{4}\omegadown\). 

The guiding centre of a magnetized particle downstream of a
superluminal, relativistic shock recedes from the shock front at a
substantial fraction of \(c\), which led \citet{2001MNRAS.328..393A}
to the assertion that Fermi acceleration would be ineffective in the
absence of strong cross-field transport (i.~e., strong scattering)
downstream. As a consequence, they presented results only for
\({\omegadown}=0\). The two limiting cases of strong and weak
scattering upstream (\({\omegaup}=0\) and \(\nuup=0\), respectively)
were investigated and yielded results which did not differ
significantly from the analytic approximation for the parallel shock
case (\({\omegaup}={\omegadown}=0\)) treated in
section~\ref{sec:pure_scattering}.

Since we take account of the energy dependence of
\(\nuupdown\), these different scattering regimes map into
different ranges of Lorentz factors for the accelerated particles.
Thus, strong and weak scattering downstream correspond to
\(\gamma\ll\gammamaxdown\) and 
\(\gamma\gg\gammamaxdown\), respectively, where
\begin{align}
  \gammamaxdown&=\nuzerodown/{\omegadown}
                 \label{eq:dmldefinition}
\end{align}
is the {\em downstream magnetized limit} to which we refer in
\citet{2022ApJ...925..182H}.
Similarly, it is possible to define a corresponding upstream limit,
taking account of the fact that a particle is scattered or
deflected through only a small angle \(\sim1/\Gamma_{\rm sh}\) whilst upstream,
before being
overtaken by the shock:
\begin{align}
  \gammamaxup&=2\sqrt{2}\Gamma_{\rm sh}\nuzeroup/{\omegadown}\,.
\label{eq:upstreammagnetizedlimit}
\end{align}

Simulations are shown in Fig~\ref{fig:spectrum_weakest}, for
\(\nuzeroup =10^{-4}\nuzerodown\), 
 implying \(\gammamaxdown = 3 \times 10^4\) and 
\(\gammamaxup\approx600\ll \gammamaxdown\).
The stationary,  power-law part of the spectrum remains close
to the analytic prediction, whereas the time-dependent cut off advances well beyond
\(\gammamaxup\).
In this case, particles in the range
\(\gammamaxup\ll\gamma\ll\gammamaxdown\) suffer deflection rather than scattering whilst upstream and the average time they spend there is
\begin{align}
  \deltatup&=2\sqrt{2}\gamma/\omegadown
             \nonumber\\
           &\approx\left(3\gammamaxdown/\gamma\right)\deltatdown
             \,.
\end{align}
Thus, they spend most of a cycle in the upstream, and the cut-off
advances with time in this range
according to \(\gammacut\approx\omegadown t/3\), i.~e., essentially at the
rate corresponding to {\em Bohm} scattering (see section~\ref{discussion}).
Nevertheless, close to \(\gammamaxdown\), eq~(\ref{eq:maxgammaestimate}) remains a good order of magnitude estimate, since here particles
divide their time almost equally between
the two regions.
At
\(\omegadown t=6\times 10^4\), \(\gammacut\) is roughly a factor of \(3\)
less than \(\gammamaxdown\),
  whereas at \(\omegadown t> 10^6\) a stationary state is 
reached at essentially all energies to which particles can be accelerated, confirming 
that saturation occurs at roughly
\(\gammamaxdown\), in agreement with the findings of
\citet{2001MNRAS.328..393A}.

\begin{figure}
\begin{center}
\includegraphics[scale=0.35]{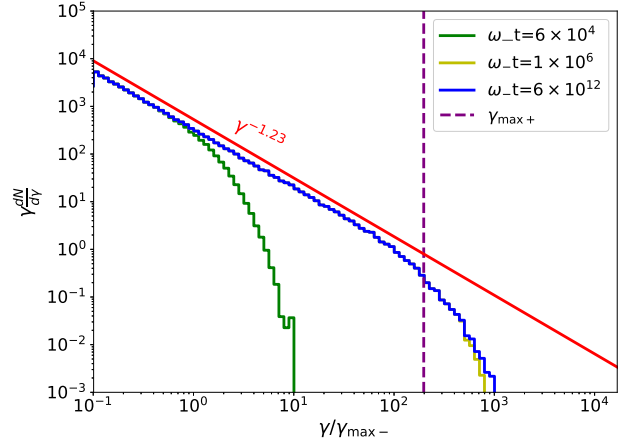}
\caption{
  Particle spectra at a perpendicular shock in a uniform magnetic field,
  with relatively strong
  scattering upstream: \(\nuzeroup\, =\, \nuzerodown\) and
  \({\nuzerodown}/\omegadown = 3 \times 10^4 = \gammamaxdown\),
  plotted as a function of \(\gamma/\gammamaxdown\).
  The vertical dashed line shows the upstream magnetized limit,
\(\gammamaxup\), see eq~(\ref{eq:upstreammagnetizedlimit}).
  \label{fig:spectrum_strongest}}
\end{center}
\end{figure}

If, however, \(\gammamaxup\gg \gammamaxdown\), the situation changes.
In Fig.~\ref{fig:spectrum_strongest} we show results for
\(\nuzeroup\, =\,  \nuzerodown\), implying
\(\gammamaxdown = 3 \times 10^4\)
and 
\(\gammamaxup=6\times10^6\gg \gammamaxdown\).
In this case the power-law spectrum does not cut off at the downstream magnetized limit, but extends up to \(\gammamaxup\).
Particles in the range \(\gammamaxdown\ll\gamma\ll\gammamaxup\)
now spend most of a cycle in the downstream region,
from which they are ejected after a fraction of a gyration:
\begin{align}
  \deltatdown&=\gamma/\omegadown
             \nonumber\\
           &\approx\left(\gammamaxup/3\gamma\right)\deltatup\,.
\end{align}
Once again, therefore, the cut-off advances linearly with time:
\(\gammacut\approx\omegadown t\), corresponding to Bohm scattering.

The stationary spectra in the regime
\(\gammamaxdown\ll\gamma\ll\gammamaxup\) 
were analyzed by \citet{Kirketal23}, using both our Monte-Carlo code and an analytic
approximation scheme. For \(\Gamma_{\rm sh}>50\) a power-law spectrum
with index \(f\propto \gamma^{-4.17}\) was found. However, the 
transition from a power-law of index \(1.23\) to one of \(1.17\) is not
discernible in Fig~\ref{fig:spectrum_strongest}, because of the relatively
small range of Lorentz factors between \(\gammamaxdown\)
and \(\gammamaxup\).

We note that in the case of an oblique shock front, it is only the component of the magnetic field perpendicular to the shock
normal in the upstream that causes particles to leave the narrow cone
in which they can move ahead of the shock front. This implies
that the \emph{upstream} magnetized limit may depend on the shock
obliquity for superluminal shocks, even though
the regular magnetic field in
the downstream remains almost perpendicular to the shock normal.
Denoting by   
\(\alpha\)
the angle between the upstream magnetic field, as measured by an
upstream observer, and the shock normal, we find
\begin{align}
  \gammamaxup&=2\sqrt{2}\Gamma_{\rm sh}\nuzeroup/\left(\left|\sin\alpha\right|{\omegadown}\right)\,,\ \textrm{for } |\sin\alpha| > 1/\Gamma_{\rm{sh}}\,,
\label{eq:upstreammagnetizedlimitgeneralized}
\end{align}
generalizing eq~(\ref{eq:upstreammagnetizedlimit}).
We have performed additional simulations that confirm this increase.

\subsection{Magnetic field with cylindrical symmetry}
\label{cylinder}

To model the situation of either a reverse shock in a magnetized jet,
or the forward shock of a jet that propagates along the rotation axis
into a medium whose magnetic structure mimics the {\em Parker wind},
we choose a simple geometry for the static magnetic field in the
upstream region:
one in which only azimuthal
field components exist in cylindrical coordinates \((\rho,\theta,x)\).
Satisfying the solenoidal field condition then requires that
\(\partial B_{\theta}/\partial \theta = 0\). To maintain a finite
current in the jet, the field should approach zero on the symmetry
axis. Therefore, we select a large-scale upstream field \(\bm{\Bup}\)
that increases linearly with cylindrical radius \(\rho\) out to a
distance \(\rho_0\), outside of which it remains constant.
As in the homogeneous case, we assume that it
is compressed upon crossing the shock front, which remains in the plane
\(x=0\) and is, therefore, also a perpendicular shock:
\begin{align}
\bm{\Bupdown}& = 
               \Bzeroupdown\bm{\hat{\theta}}
               \times\left\lbrace
\begin{array}{ll}
\left({\rho}/{\rho_{0}}\right), &\rho\, \leqslant \, \rho_{0}\\
&\\
1, & \rho_{0}  \, < \, \rho\, \leqslant \, \rho_{\rm{max}}\,.
\end{array}
\right. 
\label{Eqn:CylinderField}
\end{align}
This field corresponds to a constant axial current density within
\(\rho_0\), outside of which it falls off as \(1/\rho\).  We set a
boundary at \(\rho=\rho_{\rm{max}} \gg \rho_{\rm{0}}\), and assume
that all particles that reach it escape. In this way, \(\rho_0\)
models the thickness of the current carrying region around the axis,
which is likely to be determined by microphysical processes, whereas
\(\rho_{\rm max}\) models the macroscopic geometry of the field,
comparable, for example, to the distance of the shock from the central
object/engine.  Since times and distances are
normalized to the nonrelativistic gyro frequency and radius
corresponding to 
\(\Bzerodown\), particles with
\(\gamma\gg\rho_0\) cross the axial region without significant
deflection by the magnetic field, and particles with
\(\gamma>\rho_{\rm max}\) cannot be confined within the system,
when downstream. When these particles enter the upstream region,
their gyro radius increases by a factor \(\sim\Gamma_{\rm sh}^2\).
However, the constraints on sensitivity to the axial region and on
confinement are unchanged, because their angular distribution
is tightly beamed along the axis.

In the simulations presented here
we set \(\rho_{0}\, =\, 10^{5}\) and
\(\rho_{\rm{max}}\, =\, 10^{7}\). 
As in section
\ref{uniformmagneticfield}, we select
\(\Gamma_{\rm sh}=50\sqrt{2}\) and
\(\nuzerodown=3\times10^{4}\omegadown\), but note that
\(\omegadown\) is defined by the fiducial field \(\Bzerodown\), so that
scattering is always dominant sufficiently close to the axis \(\rho=0\).
Particles are initially
injected at \(\rho \, =\, 0\), with an isotropic angular distribution
immediately downstream of the shock front.

\begin{figure}
\begin{center}
\includegraphics[scale=0.35]{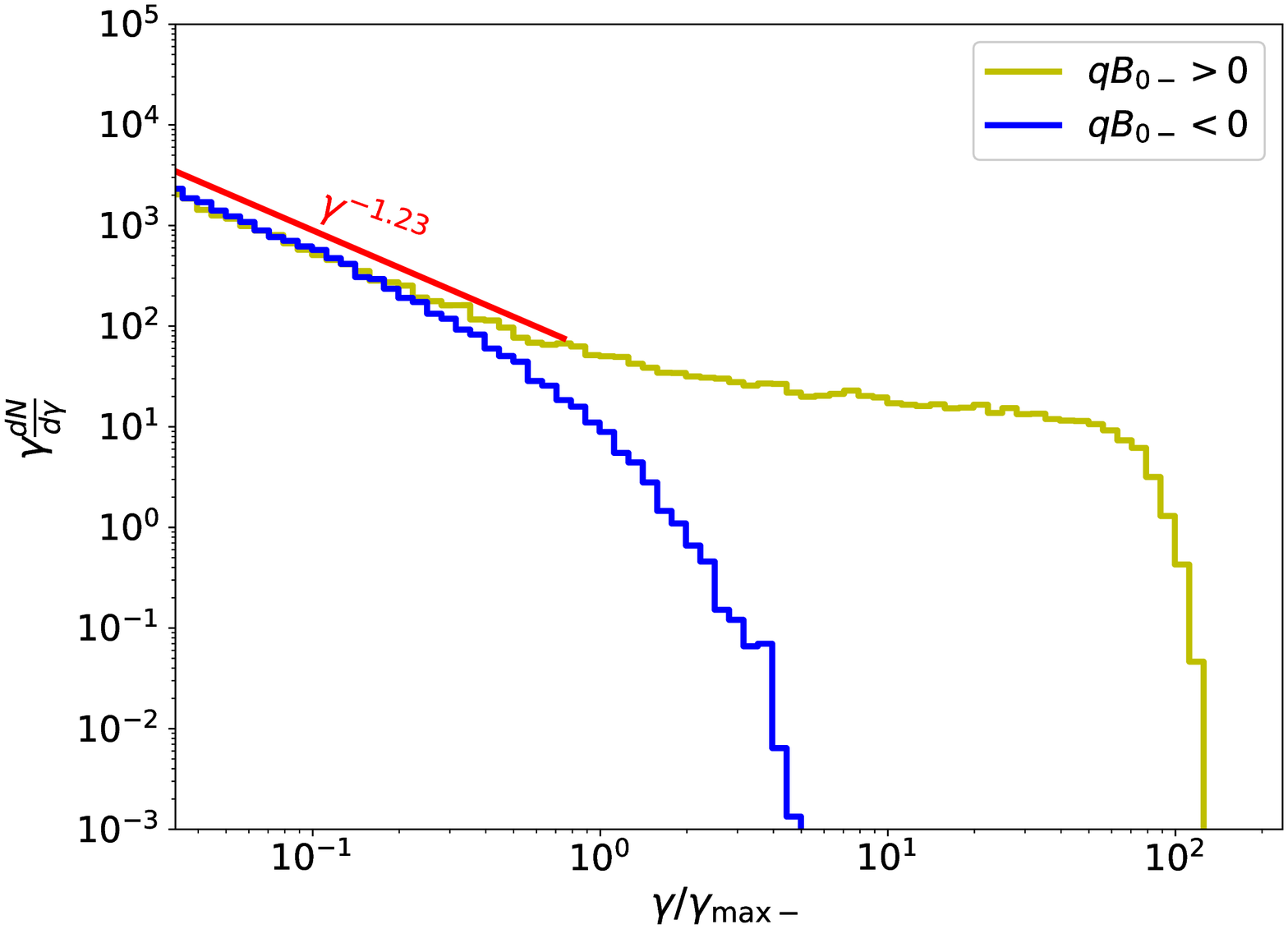}
\caption{
  Particle spectra in a cylindrically symmetric magnetic field. Here we
  consider weak scattering in the upstream, \(\nuzeroup\, =\,
  10^{-4} \nuzerodown\). The yellow and blue histograms show results
  for \(q\Bzerodown>0\) and \(q\Bzerodown<0\), respectively,
  plotted as functions of \(\gamma/\gammamaxdown\)
  (\(\gammamaxdown=3\times10^4\)). The simulations
  were run until a steady state was reached. For \(q\Bzerodown>0\) a
  cut off appears at roughly
  the confinement limit, \(\gamma\approx\rho_{\rm max}=10^7\). 
  \label{fig:cylinder}}
\end{center}
\end{figure}

The corresponding spectra for weak upstream scattering,
\(\nuzeroup \, =\, 10^{-4} \nuzerodown\), 
are shown in Fig.~\ref{fig:cylinder}.
The results shown in yellow are for test particles with the same sign of charge
as the fiducial field, \(q\Bzerodown>0\),
whereas those in blue are
for \(q\Bzerodown<0\).
These two spectra are essentially identical at low energy, but
diverge above a Lorentz factor given roughly by the downstream magnetized limit
\(\gammamaxdown\).
In the downstream region, the transport of
particles of lower Lorentz factor is everywhere dominated by scattering,
so that spectrum remains close to that expected in the parallel shock case,
or, equivalently,
in the perpendicular shock with \(\gammamaxup\ll\gammamaxdown\)
(Fig~\ref{fig:spectrum_weakest}).

At Lorentz factors greater than \(\gammamaxdown\), however, there is
a dramatic difference: the blue spectrum cuts off, whereas the yellow
spectrum hardens to higher energy until the confinement
limit \(\gamma\approx\rho_{\rm{max}}\) is reached.

\begin{figure*}
    \begin{center}
    \includegraphics[scale=0.7]{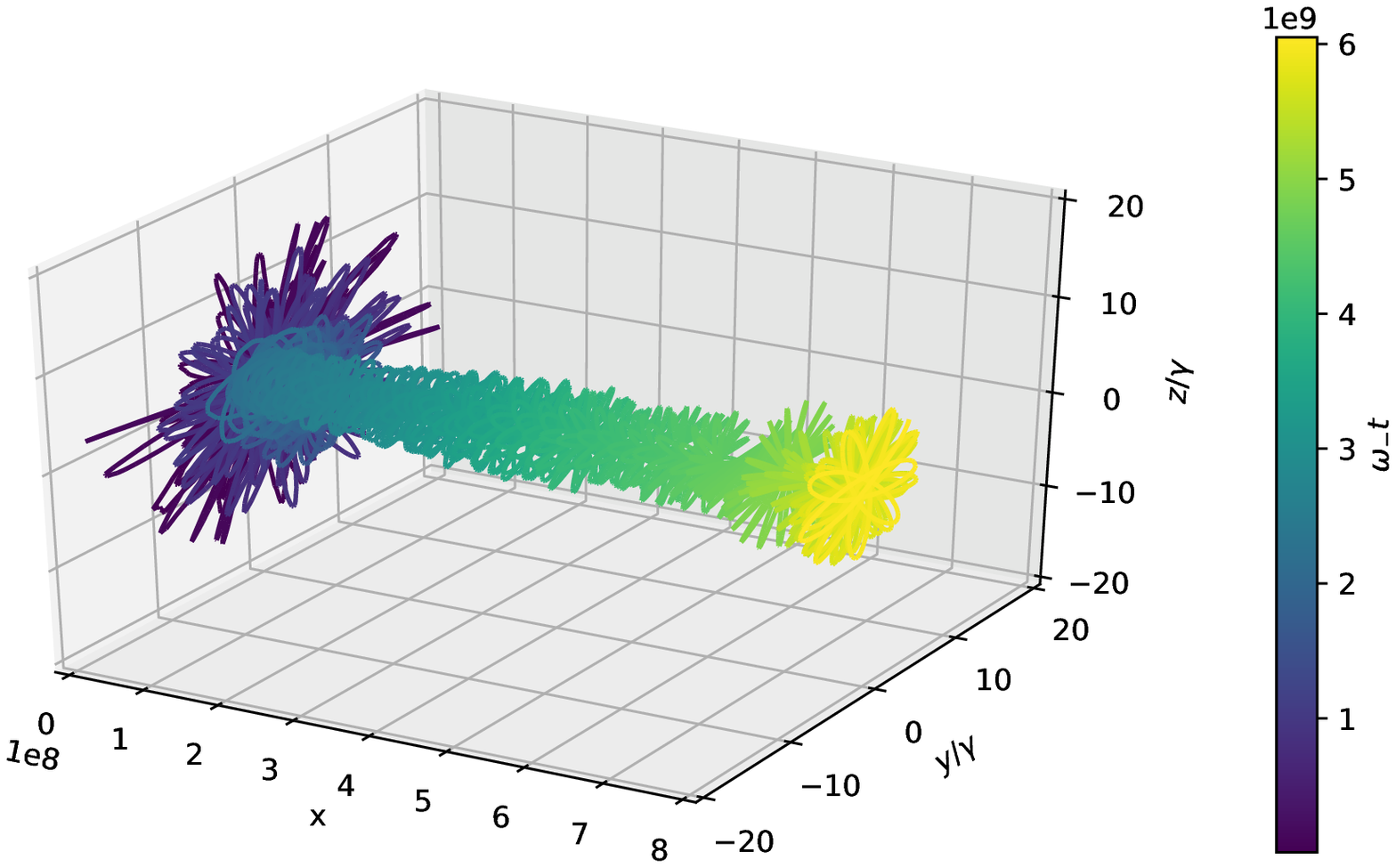}\\
    \includegraphics[scale=0.5]{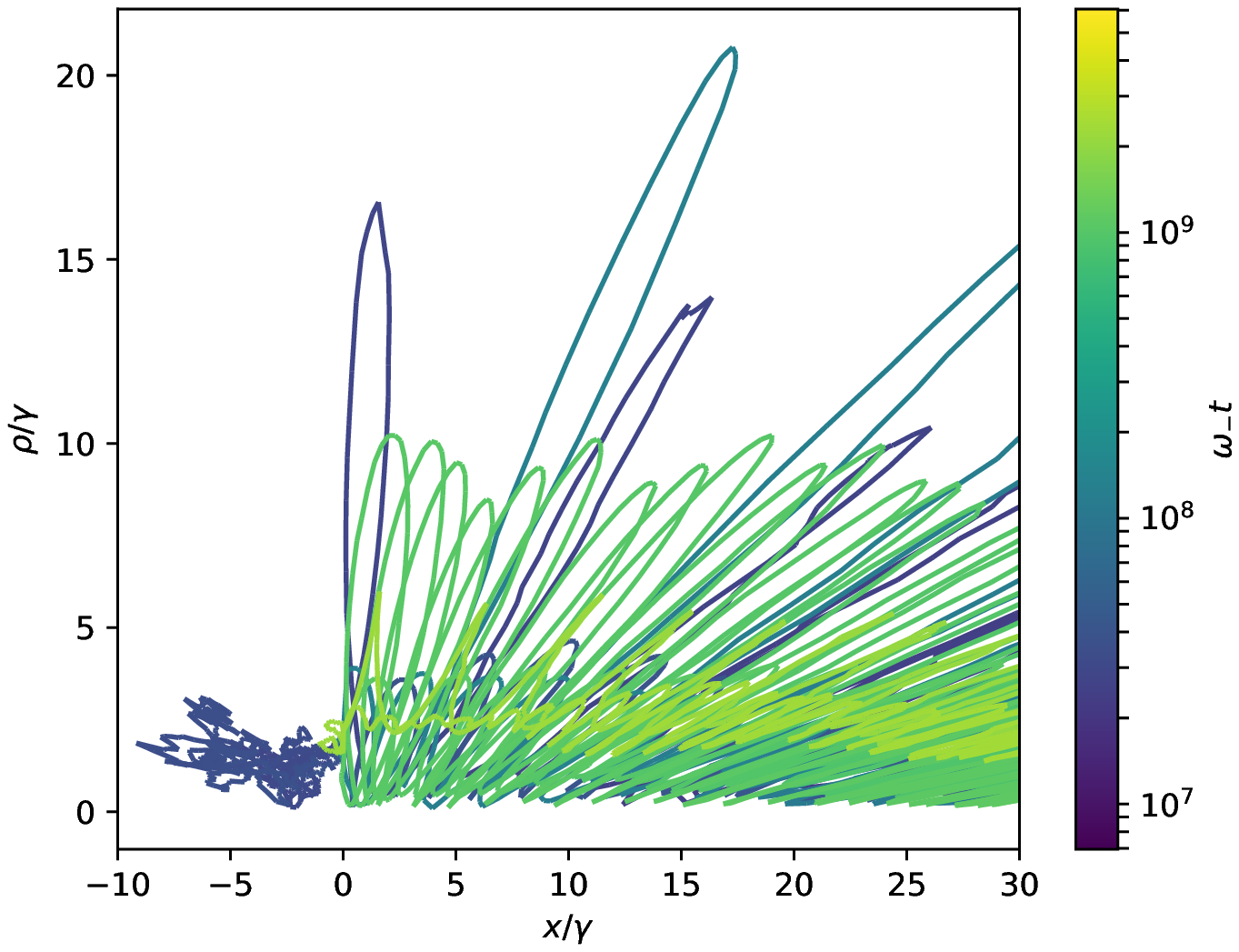}
        \includegraphics[scale=0.6]{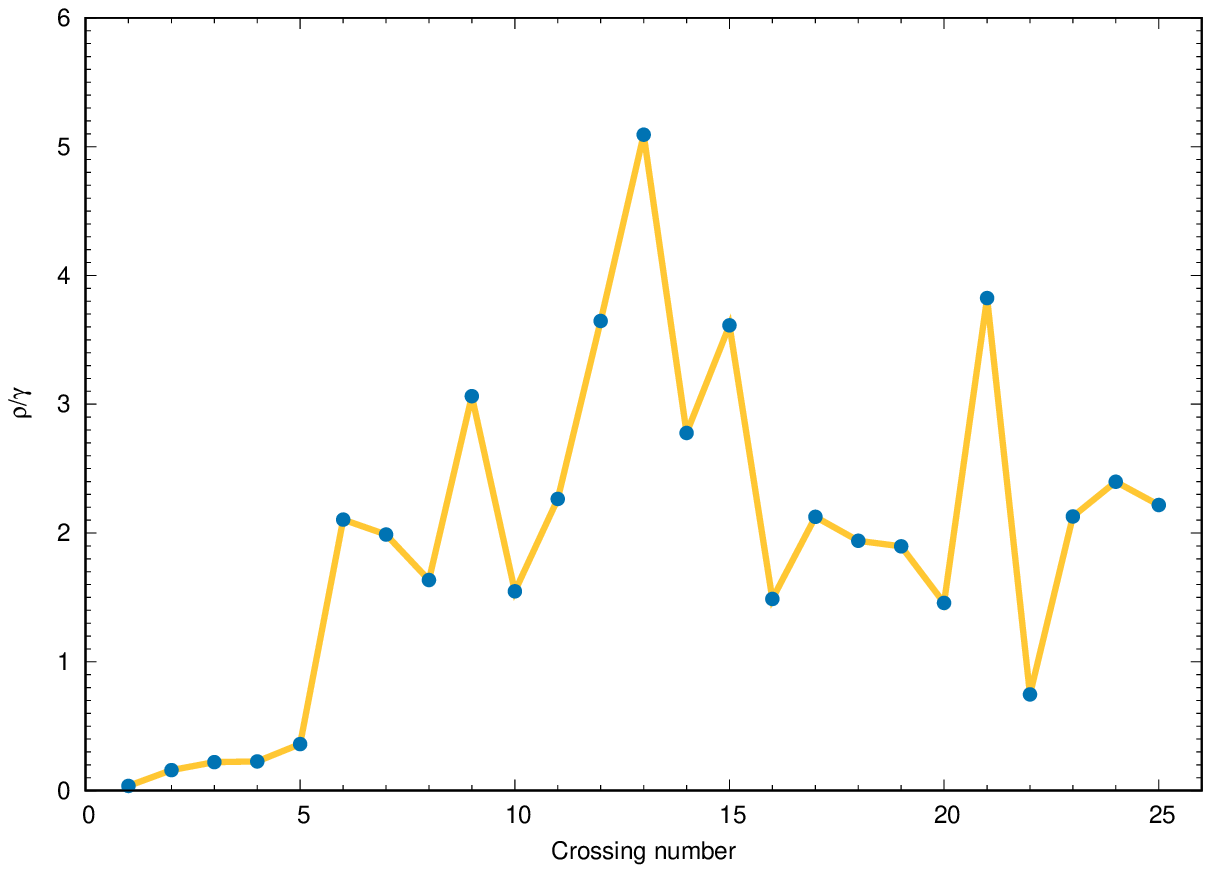}
    \caption{Trajectory in the shock rest frame of a particle
      with \(q\Bzerodown>0\) and Lorentz
      factor above \(\gammamaxdown\). Top panel: Particle
      trajectory in 3-dimensional space before escaping from the boundary
      \(\rho_{\rm{max}}\) in the upstream region.
      Bottom left panel: the trajectory in the \(x\)--\(\rho\)-plane
      near the shock front, showing a total of 25 crossings.
      Bottom right panel: The value of \(\rho/\gamma\) each time the trajectory
      crosses the shock front, showing that the trajectory is confined to
      a region close to the axis where the drift speed is \(\sim c\).
      \label{fig:trajectory}}
    \end{center}
\end{figure*}
In
a uniform field, the only drift present is the
\(\bm{E}\times\bm{B}\) drift, that causes particles to be dragged
along at the speed of the local plasma flow, independent of their
charge. In the cylindrically symmetric field, however, a
charge-dependent curvature drift is superposed on the
\(\bm{E}\times\bm{B}\)-drift. For \(qB_0>0\), this
drift is directed in the positive \(x\) direction, i.~e., in the direction of
propagation of the shock front. The drift speed is slow far from the
axis but is \(\sim c\) when \(\rho<\gamma\). As a consequence,
particles with \(q\Bzerodown>0\) that are located
close to the axis
in the downstream region and experience little scattering, are able to catch
up and overtake the shock, whereas those with \(q\Bzerodown<0\) are swept
away from it. This behaviour is illustrated
in the lower panels of
fig.~\ref{fig:trajectory}, where we show a trajectory typical
of a particle with \(q\Bzerodown>0\)
and a Lorentz factor \(>\gammamaxdown\).

\begin{figure}
\begin{center}
\includegraphics[scale=0.35]{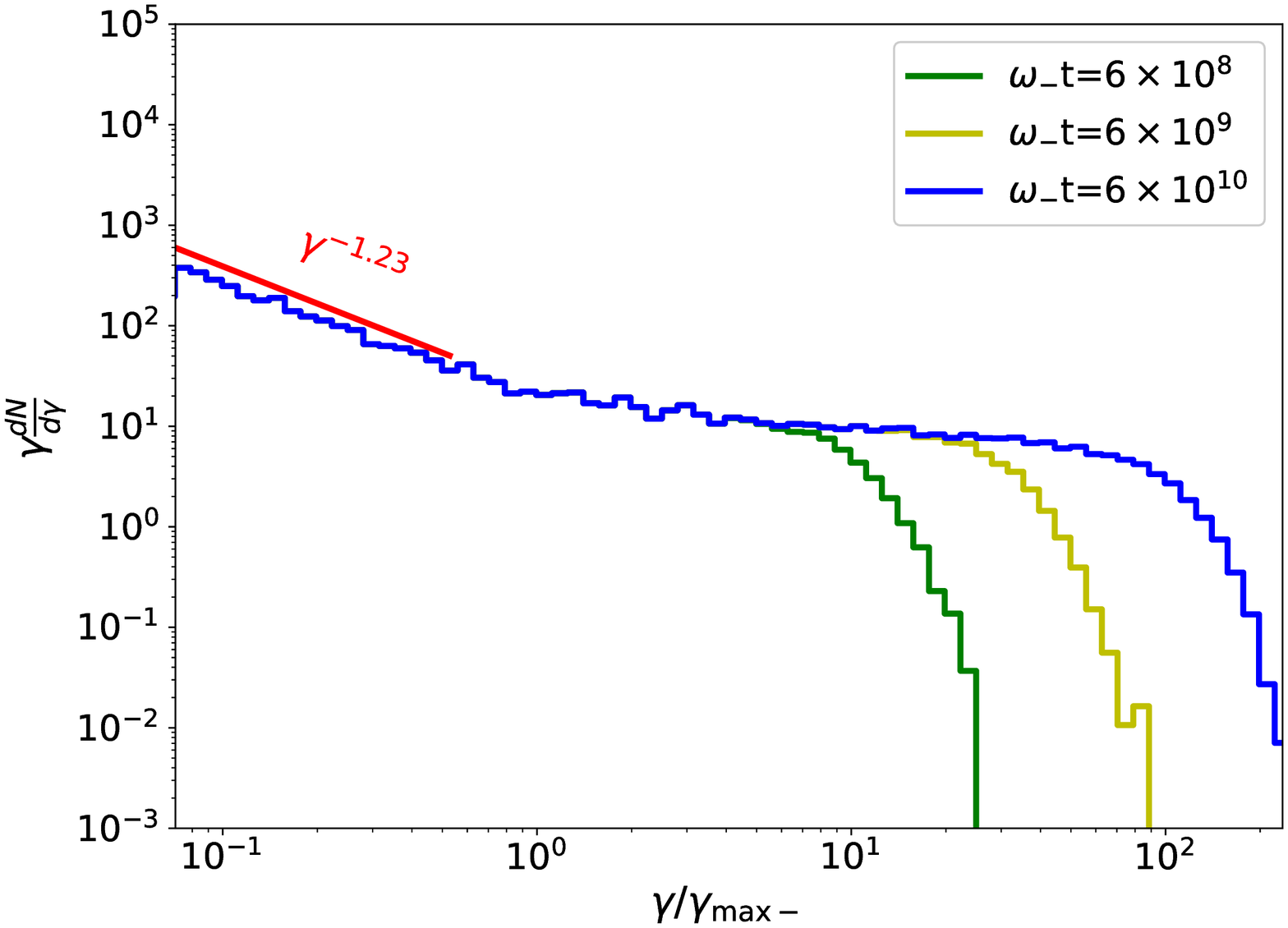}
\caption{Spectra for particles with \(q\Bzerodown>0\) and
  the same parameters as in fig~\ref{fig:cylinder},
  but without an upper limit on
  \(\rho\). The cutoff on the hard component of the
  spectrum does not saturate, but increases
  monotonically with time according to \(\gammacut\propto t^{1/2}\). 
\label{fig:cylinder_infinity}}
\end{center}
\end{figure}

The situation is reversed in
the upstream region, where particles with \(q\Bzerodown>0\)
tend to outrun
the shock front --- see the upper panel of
fig.~\ref{fig:trajectory}. However, even a very low scattering rate there
suffices eventually to deflect particles through an angle that allows
them to be caught again by the shock front. Therefore, these particles
can be accelerated beyond the limit
\(\gammamaxup\) given in eq~(\ref{eq:upstreammagnetizedlimit}), provided
the upstream region is sufficiently extended in the \(x\)-direction.
This can be seen in
Fig.~\ref{fig:cylinder_infinity}, where we run the simulation without
imposing a boundary on the radial extent of the magnetic field.
In the presence of scattering, particles can then escape only by being
swept far downstream.
In this
case, as in the case of a parallel shock in a uniform field, there is
no intrinsic upper limit on the energy to which favourably charged
particles located close to the axis can be accelerated, and the
spectrum below the time-dependent upper cut-off approaches
\(f\propto \gamma^{-3}\), as expected for particles that are
effectively confined to the vicinity of the
shock front. Because the acceleration rate
is now limited by the time spent upstream, the cut off advances
according to \(\gammacut\approx\sqrt{\Gamma_{\rm sh}\nuzeroup t}\).

\begin{figure}
\begin{center}
\includegraphics[scale=0.35]{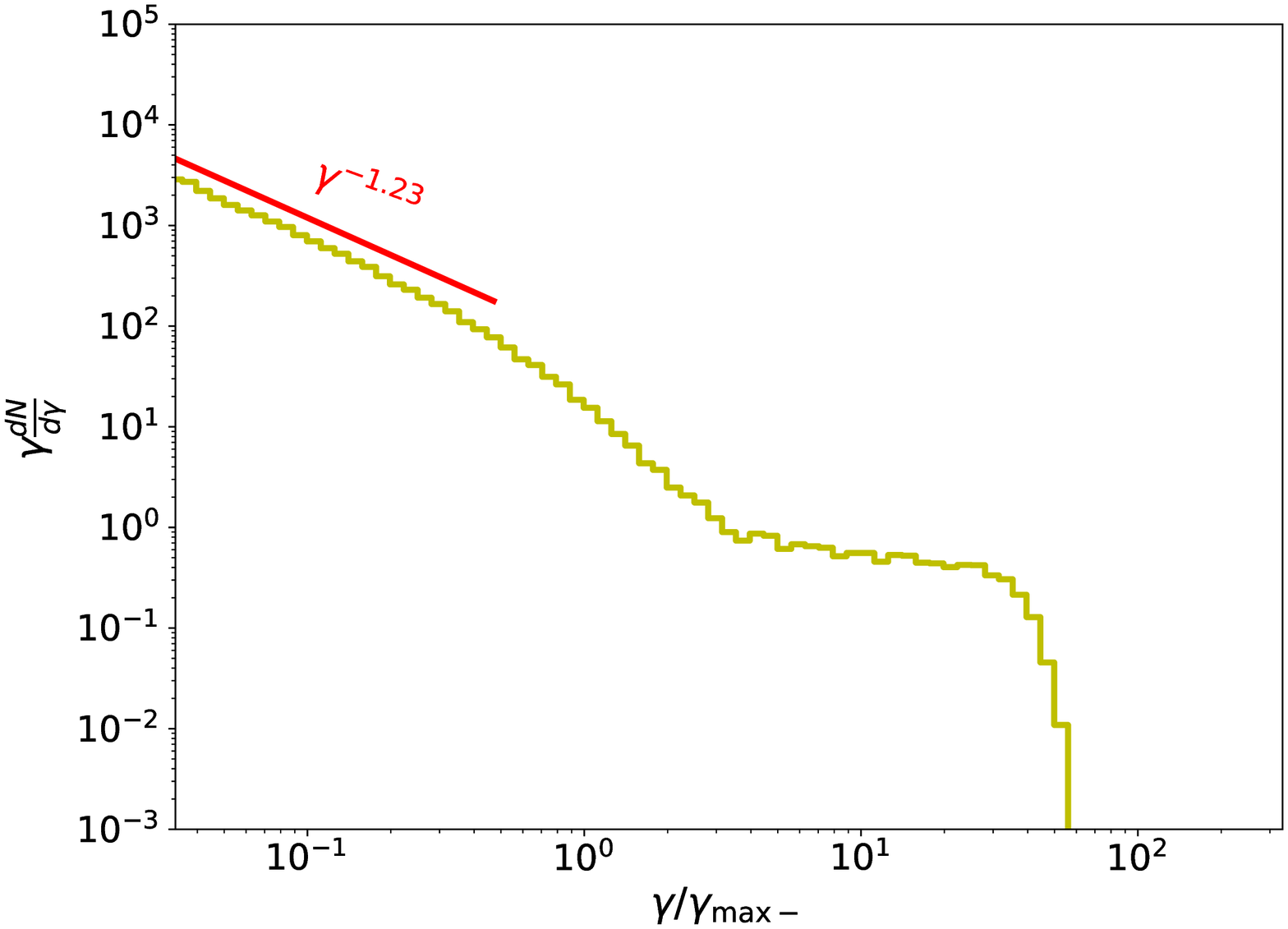}
\caption{
    The spectrum for particles with \(q\Bzerodown>0\) and
  the same parameters as in fig~\ref{fig:cylinder},
  but with an upper limit on the jet radius reduced by a factor of 2,
  i.~e., \(\rho_{\rm max}=5\times10^6\). Also, particles
  are now injected with a uniform distribution in \(\rho\)
  between the axis \(\rho = 0\) and \(\rho_{\rm{max}}\). Only those
  that reach \(\gamma\approx\gammamaxdown\) when close to the axis
  are accelerated further. 
\label{fig:cylinder1}}
\end{center}
\end{figure}

Particles far from the axis do not benefit from the curvature drift.
Hence, the positions at which particles are injected affects the
accelerated spectrum. We demonstrate this effect in
fig~\ref{fig:cylinder1}, where particles are
injected at positions that are uniformly distributed in \(\rho\)
between the axis and \(\rho_{\rm{max}}\).
Many of these particles remain far away from
the axis and populate a spectrum with index
\(f\propto\gamma^{-4.23}\) and a cut off close to \(\gammamaxdown\).
Only a small fraction of them reaches this energy whilst
sufficiently close to the axis --- i.~e., at \(\rho<\gammamaxdown\) ---
to achieve a substantial drift velocity. These are then
accelerated further
into a hard spectrum that extends up to the confinement limit.
In Fig. \ref{fig:cylinder1}, \(\rho_{\rm{max}}\) is reduced by 
a factor of 2 relative to that used in Fig.~\ref{fig:cylinder},
and the maximum energy decreases by the same factor,
thereby validating our
assertion that particles are accelerated to the confinement limit. 

\section{Discussion}
\label{discussion}

The main results of this paper concern the energy spectrum of highly
energetic charged particles that can be produced by a relativistic
shock front. They fall into two parts, each of which addresses an
idealized situation.

In the first, we consider a planar shock front that
propagates into a medium containing a relatively weak, perfectly
uniform magnetic field, upon which fluctuations are imposed.  We
assume that the uniform field lies in the plane of the shock front,
i.~e., the shock is {\em perpendicular}, since this is by far the most
likely configuration for a relativistic shock front.  The spectrum
produced then depends on the strength and length scale of the
fluctuations. Assuming these are of short length scale, as suggested
by PIC simulations, we find that there is an intrinsic limit on the
energy to which particles can be accelerated, even when other effects,
such as radiative losses, or the finite size or finite lifetime of the
system, can be neglected. This is in marked contrast to the (presumably less
realistic) situation at a parallel shock front, which does not possess
such an intrinsic limit. On the other hand, we find that the
time-asymptotic particle spectrum below this limiting energy is essentially
the same as that predicted for a parallel shock, and does not depend on the
strength of the fluctuations.

The limit itself, however, does depend on the strength of the
fluctuations in both the upstream and downstream plasma. If these are
more effective in the downstream than the upstream, we confirm the
maximum energy predicted by \citet{2022ApJ...925..182H} (the \lq\lq magnetized limit\rq\rq),
denoted here as \(\gammamaxdown\) in eq~(\ref{eq:dmldefinition}).  In
the converse situation, we confirm the spectral index indeed remains
close to that of a parallel shock, as predicted by \citet{Kirketal23}, and find a new
expression for the upper limit, \(\gammamaxup\), given in
eq~(\ref{eq:upstreammagnetizedlimit}).  In each case, we find that the
spectrum evolves with an upper cut-off energy that increases linearly
with time, as it approaches the limiting value, which is determined by
the larger of the two quantities \(\gammamaxupdown\).  This behaviour
is the same as that predicted under the assumption of {\em Bohm}
scattering, where the mean free path of a particle is assumed to equal
its gyroradius. In contrast, a time-dependence
\(\gammacut\propto t^{1/2}\) is expected when particle
transport is dominated \emph{everywhere} by nonresonant scattering \citep{KirkReville}.

The model described above is necessarily highly idealized.  For
example, the assumption of a perfectly uniform background field is
reasonable only if there are no fluctuations on a length scale
comparable to the gyroradius of the highest energy particles. This fails
if resonant interactions become important. But, more importantly
it also places a strict
lower limit \(R_{\rm up}\)
on the length scale of structures that may pre-exist in the
upstream medium. 
Although it is tempting to associate the length scale \(R_{\rm up}\) with the
overall size \(R\) of the astrophysical system under consideration, there
are several situations in which it is expected to be much smaller.
For example, MHD models of pulsar wind nebulae and jets from accreting black holes \cite[see ][for reviews]{2017SSRv..207..137P,2020ARA&A..58..407D} show both
narrow current sheets and axial current flows, where \(R_{\rm up}\ll R\).

The second idealized
situation we address is the latter: a relativistic shock propagating
along the axis of a plasma containing a cylindrically symmetric
field.
In this situation, we find that the acceleration of particles with
Lorentz 
factor \(\gamma<\gammamaxdown\) proceeds in the same manner as in a
uniform field, since their transport is dominated by scattering in the
downstream region. At higher energy, however, there is a dramatic
difference for those particles that are located within roughly
one gyroradius of the axis downstream of the shock. There, unscattered
trajectories undergo rapid curvature drifting, which enables charges of
favourable sign to catch up with the shock, thereby reducing their
probability of escaping the acceleration region. As a result, we find
a hard
\(f\propto\gamma^{-3}\) spectrum (see Fig~\ref{fig:cylinder}) for the favoured particles, that can
extend up to the confinement limit \(\gamma\approx \left|q\Bzerodown\right|R/mc^2\).
This is reminiscent of the findings of \citet{GK18}, who
integrated particle orbits directly in synthetically constructed magnetic
turbulence. They considered the equatorial region of a pulsar wind termination
shock, which contains a plane current sheet rather than an axial current, and
observed a
turbulence-dependent
hardening of the spectrum of one of the charged components, which they 
interpreted as due to {\em Speiser orbits} that cross the sheet.
In our case, the sheet is essentially contracted into the axis of the cylinder, so that it is not possible for a particle orbit to cross it. Nevertheless,
curvature drift takes over the role of the Speiser orbits and permits
particles of one charge to decouple from the
downstream fluid motion and recross into upstream. 

In interpreting our results, it is important to remember the limitations
not only of our assumed scattering model, but also of the test 
particle approximation. For example, the very hard
spectrum we find will, if it extends to high energy, begin to exert a
significant influence on the background plasma, which is not included.
In principle, PIC simulations are capable of accounting for this and other
physical effects, but they are currently challenged by
the large range of spatial and temporal scales that separate the
thermal particles from the most energetic. Thus, global
PIC simulations of both the equatorial current sheet and the axial current case
show tantalising hints of a hard spectral component
\citep{CeruttiGiacinti,CeruttiGiacinti23}, but set the
confinement limit of the simulation to
a particle Lorentz factor \(\sim 10^3\). Consequently, a distinct
hard spectral component does not emerge and the
maximum energy permitted in the simulation is well below
that expected in astrophysical objects. Similarly, 
PIC simulations of acceleration in uniform fields
report acceleration rates approximately proportional to \(t^{1/2}\)
\citep{Stockem,2013ApJ...771...54S,Plotnikovetal18}, but
did not extend to energies above the lower of the two
limits \(\gammamaxupdown\), above which we predict an unchanged spectral index
and an acceleration rate \(\propto t\).

Our principal findings on the maximum energy and spectral features have many potential applications for astrophysical sources which host relativistic shocks. Amongst the best known examples are GRBs, Pulsar wind termination shocks, and AGN/Blazars. All these source classes are well established gamma-ray emitters \cite[e.g.][]{2021ApJ...911..143A, 2021Sci...373..425L, 2021Sci...372.1081H, 2022GCN.32677....1H} and/or hard X-ray synchrotron sources \cite[e.g.][]{2018MNRAS.477.4257C,2022ApJ...941..204T}. We find that the  steady-state spectrum in almost all cases where pitch-angle diffusion dominates on at least one side of the shock is almost indiscernible from the
well-known result \(dN/d\gamma \propto \gamma^{-2.2}\), originally derived for 
exactly parallel shocks.  Although such spectra are typical of those
inferred from GRB afterglow observations, harder spectra are
not uncommon \cite[see for example][]{2018ApJ...861...85A}. Our results
indicate
that large scale field structures associated to the specific geometry
of the source, can maintain the shock acceleration cycles through the
drift motions induced by this geometry. This naturally leads to a
maximum energy close to the confinement limit, in contrast with the findings
of \citet{2018MNRAS.473.2364B}, who considered only self-excited turbulence.
It also suggests a hardening of 
the spectrum which may, in principle, be
identified with a spectral break. The asymptotic \lq
zero-escape\rq\, limit presented here; \(dN/d\gamma \propto
\gamma^{-1}\) is a result of our simplified assumptions, and we
anticipate a range of spectral shapes will be revealed in future
environment-specific studies, in particular if feedback from the axial
current on the upstream medium is included. Nevertheless, there are
several clear examples where such hard spectra are favoured. It has
been demonstrated, for example, that a
hard source spectrum for UHECR alleviates the need to invoke
problematic negative source
evolution \citep{2015PhRvD..92f3011T}. If relativistic jets, either
those of GRBs or high power AGN, are the primary source of UHECRs,
particles must be accelerated close to the confinement limit for these
sources \cite[see
  discussion in][]{2000PhST...85..191B,2020NewAR..8901543M}. Since
these particles sample the full width of the jet, they naturally probe
its large-scale underlying magnetic structure. In this regard, the
origin of the highest energy cosmic rays in our local universe is a
natural consequence of the mechanism that efficiently extracts the
power from the central engine \citep{1977MNRAS.179..433B}.

\section{Data Availability Statement}
No new data were generated or analysed in support of this research.

\bibliographystyle{mnras}
\bibliography{references}

\begin{thebibliography}{}
\makeatletter
\relax
\def\mn@urlcharsother{\let\do\@makeother \do\$\do\&\do\#\do\^\do\_\do\%\do\~}
\def\mn@doi{\begingroup\mn@urlcharsother \@ifnextchar [ {\mn@doi@}
  {\mn@doi@[]}}
\def\mn@doi@[#1]#2{\def\@tempa{#1}\ifx\@tempa\@empty \href
  {http://dx.doi.org/#2} {doi:#2}\else \href {http://dx.doi.org/#2} {#1}\fi
  \endgroup}
\def\mn@eprint#1#2{\mn@eprint@#1:#2::\@nil}
\def\mn@eprint@arXiv#1{\href {http://arxiv.org/abs/#1} {{\tt arXiv:#1}}}
\def\mn@eprint@dblp#1{\href {http://dblp.uni-trier.de/rec/bibtex/#1.xml}
  {dblp:#1}}
\def\mn@eprint@#1:#2:#3:#4\@nil{\def\@tempa {#1}\def\@tempb {#2}\def\@tempc
  {#3}\ifx \@tempc \@empty \let \@tempc \@tempb \let \@tempb \@tempa \fi \ifx
  \@tempb \@empty \def\@tempb {arXiv}\fi \@ifundefined
  {mn@eprint@\@tempb}{\@tempb:\@tempc}{\expandafter \expandafter \csname
  mn@eprint@\@tempb\endcsname \expandafter{\@tempc}}}

\bibitem[\protect\citeauthoryear{{Abdalla} et~al.,}{{Abdalla}
  et~al.}{2019}]{2019Natur.575..464A}
{Abdalla} H.,  et~al., 2019, \mn@doi [\nat] {10.1038/s41586-019-1743-9}, \href
  {https://ui.adsabs.harvard.edu/abs/2019Natur.575..464A} {575, 464}

\bibitem[\protect\citeauthoryear{{Achterberg}, {Gallant}, {Kirk}  \&
  {Guthmann}}{{Achterberg} et~al.}{2001}]{2001MNRAS.328..393A}
{Achterberg} A.,  {Gallant} Y.~A.,  {Kirk} J.~G.,   {Guthmann} A.~W.,  2001,
  \mn@doi [\mnras] {10.1046/j.1365-8711.2001.04851.x}, \href
  {https://ui.adsabs.harvard.edu/abs/2001MNRAS.328..393A} {328, 393}

\bibitem[\protect\citeauthoryear{{Ajello} et~al.,}{{Ajello}
  et~al.}{2018}]{2018ApJ...861...85A}
{Ajello} M.,  et~al., 2018, \mn@doi [\apj] {10.3847/1538-4357/aac515}, \href
  {https://ui.adsabs.harvard.edu/abs/2018ApJ...861...85A} {861, 85}

\bibitem[\protect\citeauthoryear{{Albert} et~al.,}{{Albert}
  et~al.}{2021}]{2021ApJ...911..143A}
{Albert} A.,  et~al., 2021, \mn@doi [\apj] {10.3847/1538-4357/abecda}, \href
  {https://ui.adsabs.harvard.edu/abs/2021ApJ...911..143A} {911, 143}

\bibitem[\protect\citeauthoryear{{Ballard} \& {Heavens}}{{Ballard} \&
  {Heavens}}{1991}]{1991MNRAS.251..438B}
{Ballard} K.~R.,  {Heavens} A.~F.,  1991, \mn@doi [\mnras]
  {10.1093/mnras/251.3.438}, \href
  {https://ui.adsabs.harvard.edu/abs/1991MNRAS.251..438B} {251, 438}

\bibitem[\protect\citeauthoryear{{Begelman} \& {Kirk}}{{Begelman} \&
  {Kirk}}{1990}]{1990ApJ...353...66B}
{Begelman} M.~C.,  {Kirk} J.~G.,  1990, \mn@doi [\apj] {10.1086/168590}, \href
  {https://ui.adsabs.harvard.edu/abs/1990ApJ...353...66B} {353, 66}

\bibitem[\protect\citeauthoryear{{Bell}, {Araudo}, {Matthews}  \&
  {Blundell}}{{Bell} et~al.}{2018}]{2018MNRAS.473.2364B}
{Bell} A.~R.,  {Araudo} A.~T.,  {Matthews} J.~H.,   {Blundell} K.~M.,  2018,
  \mn@doi [\mnras] {10.1093/mnras/stx2485}, \href
  {https://ui.adsabs.harvard.edu/abs/2018MNRAS.473.2364B} {473, 2364}

\bibitem[\protect\citeauthoryear{{Blandford}}{{Blandford}}{2000}]{2000PhST...85..191B}
{Blandford} R.~D.,  2000, \mn@doi [Physica Scripta Volume T]
  {10.1238/Physica.Topical.085a00191}, \href
  {https://ui.adsabs.harvard.edu/abs/2000PhST...85..191B} {85, 191}

\bibitem[\protect\citeauthoryear{{Blandford} \& {Znajek}}{{Blandford} \&
  {Znajek}}{1977}]{1977MNRAS.179..433B}
{Blandford} R.~D.,  {Znajek} R.~L.,  1977, \mn@doi [\mnras]
  {10.1093/mnras/179.3.433}, \href
  {https://ui.adsabs.harvard.edu/abs/1977MNRAS.179..433B} {179, 433}

\bibitem[\protect\citeauthoryear{{Bresci}, {Lemoine}  \& {Gremillet}}{{Bresci}
  et~al.}{2023}]{2023arXiv230311394B}
{Bresci} V.,  {Lemoine} M.,   {Gremillet} L.,  2023, \mn@doi [arXiv e-prints]
  {10.48550/arXiv.2303.11394}, \href
  {https://ui.adsabs.harvard.edu/abs/2023arXiv230311394B} {p. arXiv:2303.11394}

\bibitem[\protect\citeauthoryear{{Cerutti} \& {Giacinti}}{{Cerutti} \&
  {Giacinti}}{2020}]{CeruttiGiacinti}
{Cerutti} B.,  {Giacinti} G.,  2020, \mn@doi [\aap]
  {10.1051/0004-6361/202038883}, \href
  {https://ui.adsabs.harvard.edu/abs/2020A&A...642A.123C} {642, A123}

\bibitem[\protect\citeauthoryear{{Cerutti} \& {Giacinti}}{{Cerutti} \&
  {Giacinti}}{2023}]{CeruttiGiacinti23}
{Cerutti} B.,  {Giacinti} G.,  2023, \mn@doi [arXiv e-prints]
  {10.48550/arXiv.2303.12636}, \href
  {https://ui.adsabs.harvard.edu/abs/2023arXiv230312636C} {p. arXiv:2303.12636}

\bibitem[\protect\citeauthoryear{{Cerutti}, {Werner}, {Uzdensky}  \&
  {Begelman}}{{Cerutti} et~al.}{2013}]{2013ApJ...770..147C}
{Cerutti} B.,  {Werner} G.~R.,  {Uzdensky} D.~A.,   {Begelman} M.~C.,  2013,
  \mn@doi [\apj] {10.1088/0004-637X/770/2/147}, \href
  {https://ui.adsabs.harvard.edu/abs/2013ApJ...770..147C} {770, 147}

\bibitem[\protect\citeauthoryear{{Contopoulos} \& {Stefanou}}{{Contopoulos} \&
  {Stefanou}}{2019}]{2019MNRAS.487..952C}
{Contopoulos} I.,  {Stefanou} P.,  2019, \mn@doi [\mnras]
  {10.1093/mnras/stz1346}, \href
  {https://ui.adsabs.harvard.edu/abs/2019MNRAS.487..952C} {487, 952}

\bibitem[\protect\citeauthoryear{{Costamante}, {Bonnoli}, {Tavecchio},
  {Ghisellini}, {Tagliaferri}  \& {Khangulyan}}{{Costamante}
  et~al.}{2018}]{2018MNRAS.477.4257C}
{Costamante} L.,  {Bonnoli} G.,  {Tavecchio} F.,  {Ghisellini} G.,
  {Tagliaferri} G.,   {Khangulyan} D.,  2018, \mn@doi [\mnras]
  {10.1093/mnras/sty857}, \href
  {https://ui.adsabs.harvard.edu/abs/2018MNRAS.477.4257C} {477, 4257}

\bibitem[\protect\citeauthoryear{{Davis} \& {Tchekhovskoy}}{{Davis} \&
  {Tchekhovskoy}}{2020}]{2020ARA&A..58..407D}
{Davis} S.~W.,  {Tchekhovskoy} A.,  2020, \mn@doi [\araa]
  {10.1146/annurev-astro-081817-051905}, \href
  {https://ui.adsabs.harvard.edu/abs/2020ARA&A..58..407D} {58, 407}

\bibitem[\protect\citeauthoryear{{Giacinti} \& {Kirk}}{{Giacinti} \&
  {Kirk}}{2018}]{GK18}
{Giacinti} G.,  {Kirk} J.~G.,  2018, \mn@doi [\apj] {10.3847/1538-4357/aacffb},
  \href {https://ui.adsabs.harvard.edu/abs/2018ApJ...863...18G} {863, 18}

\bibitem[\protect\citeauthoryear{{H.~E.~S.~S. Collaboration}
  et~al.,}{{H.~E.~S.~S. Collaboration} et~al.}{2021}]{2021Sci...372.1081H}
{H.~E.~S.~S. Collaboration} et~al., 2021, \mn@doi [Science]
  {10.1126/science.abe8560}, \href
  {https://ui.adsabs.harvard.edu/abs/2021Sci...372.1081H} {372, 1081}

\bibitem[\protect\citeauthoryear{{Huang}, {Kirk}, {Giacinti}  \&
  {Reville}}{{Huang} et~al.}{2022a}]{2022ApJ...925..182H}
{Huang} Z.-Q.,  {Kirk} J.~G.,  {Giacinti} G.,   {Reville} B.,  2022a, \mn@doi
  [\apj] {10.3847/1538-4357/ac3f38}, \href
  {https://ui.adsabs.harvard.edu/abs/2022ApJ...925..182H} {925, 182}

\bibitem[\protect\citeauthoryear{{Huang}, {Hu}, {Chen}, {Zha}, {Liu}, {Yao},
  {Cao}  \& {Experiment}}{{Huang} et~al.}{2022b}]{2022GCN.32677....1H}
{Huang} Y.,  {Hu} S.,  {Chen} S.,  {Zha} M.,  {Liu} C.,  {Yao} Z.,  {Cao} Z.,
  {Experiment} T.~L.,  2022b, GRB Coordinates Network, \href
  {https://ui.adsabs.harvard.edu/abs/2022GCN.32677....1H} {32677, 1}

\bibitem[\protect\citeauthoryear{{Kirk} \& {Duffy}}{{Kirk} \&
  {Duffy}}{1999}]{1999JPhG...25R.163K}
{Kirk} J.~G.,  {Duffy} P.,  1999, \mn@doi [Journal of Physics G Nuclear
  Physics] {10.1088/0954-3899/25/8/201}, \href
  {https://ui.adsabs.harvard.edu/abs/1999JPhG...25R.163K} {25, R163}

\bibitem[\protect\citeauthoryear{{Kirk} \& {Reville}}{{Kirk} \&
  {Reville}}{2010}]{KirkReville}
{Kirk} J.~G.,  {Reville} B.,  2010, \mn@doi [\apjl]
  {10.1088/2041-8205/710/1/L16}, \href
  {https://ui.adsabs.harvard.edu/abs/2010ApJ...710L..16K} {710, L16}

\bibitem[\protect\citeauthoryear{{Kirk} \& {Schneider}}{{Kirk} \&
  {Schneider}}{1988}]{1988A&A...201..177K}
{Kirk} J.~G.,  {Schneider} P.,  1988, \aap, \href
  {https://ui.adsabs.harvard.edu/abs/1988A&A...201..177K} {201, 177}

\bibitem[\protect\citeauthoryear{{Kirk}, {Guthmann}, {Gallant}  \&
  {Achterberg}}{{Kirk} et~al.}{2000}]{2000ApJ...542..235K}
{Kirk} J.~G.,  {Guthmann} A.~W.,  {Gallant} Y.~A.,   {Achterberg} A.,  2000,
  \mn@doi [\apj] {10.1086/309533}, \href
  {https://ui.adsabs.harvard.edu/abs/2000ApJ...542..235K} {542, 235}

\bibitem[\protect\citeauthoryear{{Kirk}, {Reville}  \& {Huang}}{{Kirk}
  et~al.}{2023}]{Kirketal23}
{Kirk} J.~G.,  {Reville} B.,   {Huang} Z.-Q.,  2023, \mn@doi [\mnras]
  {10.1093/mnras/stac3589}, \href
  {https://ui.adsabs.harvard.edu/abs/2023MNRAS.519.1022K} {519, 1022}

\bibitem[\protect\citeauthoryear{{Lemoine} \& {Revenu}}{{Lemoine} \&
  {Revenu}}{2006}]{2006MNRAS.366..635L}
{Lemoine} M.,  {Revenu} B.,  2006, \mn@doi [\mnras]
  {10.1111/j.1365-2966.2005.09912.x}, \href
  {https://ui.adsabs.harvard.edu/abs/2006MNRAS.366..635L} {366, 635}

\bibitem[\protect\citeauthoryear{{Lhaaso Collaboration} et~al.,}{{Lhaaso
  Collaboration} et~al.}{2021}]{2021Sci...373..425L}
{Lhaaso Collaboration} et~al., 2021, \mn@doi [Science]
  {10.1126/science.abg5137}, \href
  {https://ui.adsabs.harvard.edu/abs/2021Sci...373..425L} {373, 425}

\bibitem[\protect\citeauthoryear{{MAGIC Collaboration} et~al.,}{{MAGIC
  Collaboration} et~al.}{2019}]{2019Natur.575..459M}
{MAGIC Collaboration} et~al., 2019, \mn@doi [\nat] {10.1038/s41586-019-1754-6},
  \href {https://ui.adsabs.harvard.edu/abs/2019Natur.575..459M} {575, 459}

\bibitem[\protect\citeauthoryear{{Matthews}, {Bell}  \& {Blundell}}{{Matthews}
  et~al.}{2020}]{2020NewAR..8901543M}
{Matthews} J.~H.,  {Bell} A.~R.,   {Blundell} K.~M.,  2020, \mn@doi [\nar]
  {10.1016/j.newar.2020.101543}, \href
  {https://ui.adsabs.harvard.edu/abs/2020NewAR..8901543M} {89, 101543}

\bibitem[\protect\citeauthoryear{{Medvedev} \& {Zakutnyaya}}{{Medvedev} \&
  {Zakutnyaya}}{2009}]{2009ApJ...696.2269M}
{Medvedev} M.~V.,  {Zakutnyaya} O.~V.,  2009, \mn@doi [\apj]
  {10.1088/0004-637X/696/2/2269}, \href
  {https://ui.adsabs.harvard.edu/abs/2009ApJ...696.2269M} {696, 2269}

\bibitem[\protect\citeauthoryear{{Milosavljevi{\'c}} \&
  {Nakar}}{{Milosavljevi{\'c}} \& {Nakar}}{2006}]{2006ApJ...651..979M}
{Milosavljevi{\'c}} M.,  {Nakar} E.,  2006, \mn@doi [\apj] {10.1086/507975},
  \href {https://ui.adsabs.harvard.edu/abs/2006ApJ...651..979M} {651, 979}

\bibitem[\protect\citeauthoryear{{Niemiec}, {Ostrowski}  \& {Pohl}}{{Niemiec}
  et~al.}{2006}]{2006ApJ...650.1020N}
{Niemiec} J.,  {Ostrowski} M.,   {Pohl} M.,  2006, \mn@doi [\apj]
  {10.1086/506901}, \href
  {https://ui.adsabs.harvard.edu/abs/2006ApJ...650.1020N} {650, 1020}

\bibitem[\protect\citeauthoryear{{Ostrowski}}{{Ostrowski}}{1993}]{1993MNRAS.264..248O}
{Ostrowski} M.,  1993, \mn@doi [\mnras] {10.1093/mnras/264.1.248}, \href
  {https://ui.adsabs.harvard.edu/abs/1993MNRAS.264..248O} {264, 248}

\bibitem[\protect\citeauthoryear{{Plotnikov}, {Grassi}  \& {Grech}}{{Plotnikov}
  et~al.}{2018}]{Plotnikovetal18}
{Plotnikov} I.,  {Grassi} A.,   {Grech} M.,  2018, \mn@doi [\mnras]
  {10.1093/mnras/sty979}, \href
  {https://ui.adsabs.harvard.edu/abs/2018MNRAS.477.5238P} {477, 5238}

\bibitem[\protect\citeauthoryear{{Porth}, {Buehler}, {Olmi}, {Komissarov},
  {Lamberts}, {Amato}, {Yuan}  \& {Rudy}}{{Porth}
  et~al.}{2017}]{2017SSRv..207..137P}
{Porth} O.,  {Buehler} R.,  {Olmi} B.,  {Komissarov} S.,  {Lamberts} A.,
  {Amato} E.,  {Yuan} Y.,   {Rudy} A.,  2017, \mn@doi [\ssr]
  {10.1007/s11214-017-0344-x}, \href
  {https://ui.adsabs.harvard.edu/abs/2017SSRv..207..137P} {207, 137}

\bibitem[\protect\citeauthoryear{{Press}, {Teukolsky}, {Vetterling}  \&
  {Flannery}}{{Press} et~al.}{1992}]{NumericalRecipes}
{Press} W.~H.,  {Teukolsky} S.~A.,  {Vetterling} W.~T.,   {Flannery} B.~P.,
  1992, {Numerical recipes in FORTRAN. The art of scientific computing}

\bibitem[\protect\citeauthoryear{{Reville} \& {Bell}}{{Reville} \&
  {Bell}}{2014}]{2014MNRAS.439.2050R}
{Reville} B.,  {Bell} A.~R.,  2014, \mn@doi [\mnras] {10.1093/mnras/stu088},
  \href {https://ui.adsabs.harvard.edu/abs/2014MNRAS.439.2050R} {439, 2050}

\bibitem[\protect\citeauthoryear{{Sironi}, {Spitkovsky}  \& {Arons}}{{Sironi}
  et~al.}{2013}]{2013ApJ...771...54S}
{Sironi} L.,  {Spitkovsky} A.,   {Arons} J.,  2013, \mn@doi [\apj]
  {10.1088/0004-637X/771/1/54}, \href
  {https://ui.adsabs.harvard.edu/abs/2013ApJ...771...54S} {771, 54}

\bibitem[\protect\citeauthoryear{{Sironi}, {Plotnikov}, {N{\"a}ttil{\"a}}  \&
  {Beloborodov}}{{Sironi} et~al.}{2021}]{2021PhRvL.127c5101S}
{Sironi} L.,  {Plotnikov} I.,  {N{\"a}ttil{\"a}} J.,   {Beloborodov} A.~M.,
  2021, \mn@doi [\prl] {10.1103/PhysRevLett.127.035101}, \href
  {https://ui.adsabs.harvard.edu/abs/2021PhRvL.127c5101S} {127, 035101}

\bibitem[\protect\citeauthoryear{{Stockem}, {Fi{\'u}za}, {Fonseca}  \&
  {Silva}}{{Stockem} et~al.}{2012}]{Stockem}
{Stockem} A.,  {Fi{\'u}za} F.,  {Fonseca} R.~A.,   {Silva} L.~O.,  2012,
  \mn@doi [\apj] {10.1088/0004-637X/755/1/68}, \href
  {https://ui.adsabs.harvard.edu/abs/2012ApJ...755...68S} {755, 68}

\bibitem[\protect\citeauthoryear{{Summerlin} \& {Baring}}{{Summerlin} \&
  {Baring}}{2012}]{2012ApJ...745...63S}
{Summerlin} E.~J.,  {Baring} M.~G.,  2012, \mn@doi [\apj]
  {10.1088/0004-637X/745/1/63}, \href
  {https://ui.adsabs.harvard.edu/abs/2012ApJ...745...63S} {745, 63}

\bibitem[\protect\citeauthoryear{{Takamoto} \& {Kirk}}{{Takamoto} \&
  {Kirk}}{2015}]{2015ApJ...809...29T}
{Takamoto} M.,  {Kirk} J.~G.,  2015, \mn@doi [\apj]
  {10.1088/0004-637X/809/1/29}, \href
  {https://ui.adsabs.harvard.edu/abs/2015ApJ...809...29T} {809, 29}

\bibitem[\protect\citeauthoryear{{Taylor}, {Ahlers}  \& {Hooper}}{{Taylor}
  et~al.}{2015}]{2015PhRvD..92f3011T}
{Taylor} A.~M.,  {Ahlers} M.,   {Hooper} D.,  2015, \mn@doi [\prd]
  {10.1103/PhysRevD.92.063011}, \href
  {https://ui.adsabs.harvard.edu/abs/2015PhRvD..92f3011T} {92, 063011}

\bibitem[\protect\citeauthoryear{{Thimmappa}, {Stawarz}, {Neilsen}, {Ostrowski}
   \& {Reville}}{{Thimmappa} et~al.}{2022}]{2022ApJ...941..204T}
{Thimmappa} R.,  {Stawarz} {\L}.,  {Neilsen} J.,  {Ostrowski} M.,   {Reville}
  B.,  2022, \mn@doi [\apj] {10.3847/1538-4357/aca472}, \href
  {https://ui.adsabs.harvard.edu/abs/2022ApJ...941..204T} {941, 204}

\bibitem[\protect\citeauthoryear{{Vanthieghem}, {Lemoine}, {Plotnikov},
  {Grassi}, {Grech}, {Gremillet}  \& {Pelletier}}{{Vanthieghem}
  et~al.}{2020}]{2020Galax...8...33V}
{Vanthieghem} A.,  {Lemoine} M.,  {Plotnikov} I.,  {Grassi} A.,  {Grech} M.,
  {Gremillet} L.,   {Pelletier} G.,  2020, \mn@doi [Galaxies]
  {10.3390/galaxies8020033}, \href
  {https://ui.adsabs.harvard.edu/abs/2020Galax...8...33V} {8, 33}

\makeatother
\end{thebibliography}

\appendix
\section{Monte-Carlo code}
\label{appendixA}
In our Monte-Carlo implementation, we update individual trajectories
over a time \(\Delta t\) using a
5th~order Runge-Kutta integrator \citep{NumericalRecipes}
with an adaptive time step, and choose
\(\Delta t\) such that the
scattering is well resolved:
Since the angular distribution function is concentrated in a
cone of opening angle \(1/\Gamma_{\rm sh}\) upstream, the condition
we use there is \(\Delta t=10^{-3}/\left(\Gamma_{\rm sh}^2\nuup\right)\).
Downstream, on the other hand, the distribution function is expected to
be a smoothly varying function of angle, so that
\(\Delta t=10^{-3}/\nudown\).
After each step \(\Delta t\),
the direction of the momentum vector changed to account for
small angle scattering, following the method described by
\cite{1988A&A...201..177K}. If a particle crosses the shock front
during \(\Delta t\), a root finding algorithm is applied to truncate
this step such that it ends precisely on the shock surface. A Lorentz
transformation is then made to the new frame and a new step is taken
in this frame before applying the next scattering.  Since we consider
relativistic shocks, where the probability of a particle returning to the
shock front after entering the downstream region is relatively low
(\(\sim50\%\)) a particle splitting method is adopted. Each particle
is initially assigned a weight of unity. On every third shock crossing
from upstream to downstream, eight daughter particles are created,
with a weight adjusted accordingly, and are then followed along
statistically independent paths. We set an upper limit of ten on the
number of generations of daughter particles.

At \(t=0\), particle trajectories are initiated isotropically
immediately downstream of the
shock with a Lorentz factor twice that of the shock.
The energy spectrum,
and angular distribution are found by recording the momentum \(\bm{p}\)
each time a trajectory or one of its daughters
crosses the shock front, until either the time elapsed, \(t\) 
(measured in the downstream), reaches
a pre-determined value, or the particle moves sufficiently far
from the shock downstream. We find the results are insensitive to this boundary
provided it exceeds either 10~times the gyro-radius \(\gamma c/{\omegadown}\)
or 100 times the scattering mean free path \(c/\nu_{\rm d}\).
In this way, a time-dependent solution of Eq.~(\ref{fpeq}) is simulated
with an injection term that is zero for \(t<0\) and constant for \(t>0\).

\end{document}